\documentclass[aps,prr,twocolumn,reprint,superscriptaddress,longbibliography]{revtex4-2}

\usepackage{hyperref}
\usepackage{graphicx}  % needed for figures
\usepackage{bm}        % for math
\usepackage{amssymb}   % for math
\usepackage{xspace}
\usepackage{verbatim}
\usepackage{color}
\usepackage{soul}
\usepackage{subfigure}
\usepackage[export]{adjustbox}
\usepackage{dcolumn}
\usepackage{amsthm,stmaryrd,mathtools}
\usepackage{txfonts}
\usepackage{braket}
\usepackage{amsmath}
\usepackage{xcolor}
\usepackage{array}
\usepackage{multirow}
\usepackage{tabularx}
\usepackage{booktabs} 
\usepackage{lipsum}
\usepackage{dsfont}
\definecolor{lightblue}{RGB}{100,100,250}

\begin{document}

\global\long\def\ket#1{\left|#1\right\rangle }%
\global\long\def\bra#1{\left\langle #1\right|}%
\global\long\def\braket#1#2{\langle#1|#2\rangle}%
\global\long\def\expectation#1#2#3{\langle#1|#2|#3\rangle}%
\global\long\def\average#1{\langle#1\rangle}%

\author{Rozhin Yousefjani}%
\email{ryousefjani@hbku.edu.qa}
\affiliation{Qatar Center for Quantum Computing, College of Science and Engineering, Hamad Bin Khalifa University, Doha, Qatar}

\author{Saif Al-Kuwari}%
\affiliation{Qatar Center for Quantum Computing, College of Science and Engineering, Hamad Bin Khalifa University, Doha, Qatar} 
\author{Abolfazl Bayat}%
\affiliation{Institute of Fundamental and Frontier Sciences, University of Electronic Science and Technology of China, Chengdu 610051, China}
\affiliation{Key Laboratory of Quantum Physics and Photonic Quantum Information, Ministry of Education, University of Electronic Science
and Technology of China, Chengdu 611731, China}

% \title{Discrete time crystal for quantum-enhanced sensing of periodic-fields}
\title{Discrete time crystal for periodic-field sensing with quantum-enhanced precision}

\begin{abstract}
Sensing periodic-fields using quantum sensors has been an active field of research. In many of these scenarios, the quantum state of the probe is flipped regularly by the application of $\pi$-pulses to accumulate information about the target periodic-field. The emergence of a discrete time crystalline phase, as a nonequilibrium phase of matter, naturally provides oscillations in a many-body system with an inherent controllable frequency. They benefit from long coherence time and robustness against imperfections, which makes them excellent potential quantum sensors.  
In this paper, through theoretical and numerical analysis, we show that a disorder-free discrete time crystal probe can reach the ultimate achievable precision for sensing a periodic-field. As the amplitude of the periodic-field increases, the discrete time crystalline order diminishes, and the performance of the probe decreases remarkably. Nevertheless, the obtained quantum enhancement in the discrete time crystal phase, which is experimentally accessible using standard projective measurements,
shows robustness against different imperfections and dephasing noise in the protocol. Finally, we propose the implementation of our protocol in ultra-cold atoms in optical lattices.     
\end{abstract}

\maketitle

\section{Introduction }

Metrological tasks can harness quantum features, such as superposition and entanglement, to achieve greater precision compared to their classical counterparts~\cite{degen2017quantum}. 
To estimate an unknown parameter $h_a$, the precision of the estimation $\delta h_a$, quantified by standard deviation, is fundamentally lower bounded by Cram\'{e}r-Rao inequality~\cite{nla.cat-vn81100, Rao1992} as $\delta h_a{\geq}1/\sqrt{M\mathcal{F}_Q}$ where $M$ is the number of observations and $\mathcal{F}_Q$ is the quantum Fisher information (QFI)~\cite{fisher1922mathematical}. 
If the Hamiltonian $H{=}H_0{+}h_a H_1$, in which $H_{0}$ independently describes the physics of the probe and $H_{1}$ is related to the parameter, encodes $h_a$ through the unitary evolution $U(t){=}e^{-itH}$,
it has been shown that the QFI is upper bounded by $\mathcal{F}_{Q}{\leq}t^2 ||H_{1}||^2$, where $||H_{1}||$ is the spectral norm of $H_1$ and can be related to the size of the probe $L$ as $||H_{1}||^2{\propto}L^{\beta}$~\cite{boixo2007generalized,abiuso2025fundamental,puig2024dynamical}.
Although in the absence of quantum features, one obtains $\beta{=}1$, known as the standard quantum limit, their presence allows quantum-enhanced sensitivity, namely $\beta{>}1$. 
For $H_0{=}0$, this quantum enhancement can be achieved in phase shift detection by initializing the probe in an entangled Greenberger–Horne–Zeilinger (GHZ) state~\cite{giovannetti2004quantum, leibfried2004toward,giovannetti2006quantum,banaszek2009quantum,giovannetti2011advances,frowis2011stable,demkowicz2012elusive,wang2018entanglement,kwon2019nonclassicality,toth2014quantum}.
However, this approach is limited by the difficulty in creating and maintaining such GHZ-type states, especially on a large scale~\cite{yousefjani2017estimating,demkowiczchapter}. In or out of equilibrium, strongly correlated many-body systems provide a valid alternative to this approach~\cite{zanardi2006ground,zanardi2007mixed,gu2008fidelity,zanardi2008quantum,invernizzi2008optimal,gu2010fidelity,gammelmark2011phase,skotiniotis2015quantum,rams2018limits,wei2019fidelity,chu2021dynamic,liu2021experimental,montenegro2021global,mirkhalaf2021criticality,di2021critical,montenegro2024review,Rozha3,Rozha4,Rozha5}. Due to $H_0{\neq}0$, many-body sensors benefit from the entanglement that is either intrinsically present in the spectrum or generated freely during the time evolution ~\cite{montenegro2024review}.

Quantum sensors have also been developed for periodic field sensing inferring the amplitude~\cite{timoney2011quantum,baumgart2016ultrasensitive,weidt2016trapped,petrovnin2024microwave}, the frequency~\cite{lang2015dynamical,khodjasteh2005fault} and the phase~\cite{paris2009quantum,shor1994algorithms,hall2009sensing,de2011single,schoenfeld2011real,rondin2014magnetometry} of oscillatory fields in the environment.  In many such schemes, high-frequency spin echo pulses, typically used in dynamic decoupling~\cite{viola1998dynamical,viola1999dynamical,viola1999dynamical,viola1999universal,khodjasteh2005fault,bonizzoni2024quantum,degen2017quantum} to overcome decoherence, are used to improve the accumulation of information about the time-varying signal in the quantum state of the probe. However, these high-frequency pulses often diminish spin correlations in the quantum state of the probe, which may decrease its sensing capability. Recent studies demonstrate that the spin correlations created by the system's dynamics can be utilized to achieve enhanced periodic-field sensing~\cite{zhou2020quantum,mishra2022integrable}. 
Among these proposals, systems with long-range spatiotemporal ordering, known as time crystals, are attracting increasing interest~\cite{montenegro2023quantum,Iemini2023,shukla2025prethermal,lyu2020eternal,cabot2024continuous,yousefjani2025DTC}.

% {\color{red}(I THINK IN THIS PARAGRAPH YOU CAN DEFINE THE ULTIMATE BOUNDS FOR THE QFI GIVEN BY MARTI, WHICH WE CAN ALSO CALL STANDARD AND HEISENBERG LIMIT, AND DISCUSS THE IMPORTANCE OF ACHIEVING SUCH BOUNDS).}
% Conventional crystals are structures which formed by breaking spatial symmetry.  

Breaking time-translational symmetry in periodically driven non-equilibrium systems may create a new phase of matter, known as time crystalline phase~\cite{wilczek2012quantum,Bruno2013b,watanabe2015absence,Kozin2019,Sacha2015,khemani2016phase,else2016floquet}. In recent years, discrete time crystals (DTCs), have been the subject of intensive  theoretical~\cite{Sacha2015,khemani2016phase,else2016floquet,yao2017discrete,russomanno2017floquet,ho2017critical,huang2018clean,Matus2019,kshetrimayum2020stark,estarellas2020simulating,Maskara2021Discrete,Wang2021,
pizzi2021higher,Collura2022Discrete,Huang2022Discrete,
Bull2022Tuning,Deng2023Using,liu2023discrete,Huang2023Analytical,
Giergiel2023,sacha2017time,else2020discrete,khemani2019brief,SachaTC2020,Hannaford2022,Zaletel2023,chen2024discrete,marripour2025time} and experimental~\cite{zhang2017observation,
choi2017observation,pal2018temporal,rovny2018observation,
smits2018observation,Randall2021,Kessler2021,xu2021realizing,Kyprianidis2021,Taheri2022,mi2022time,frey2022realization,
Bao2024,Kazuya2024,Liu2024,Liu2024a,moon2024discrete} investigations. 
In periodically driven systems with a period $T$, DTCs 
are characterized by subharmonic responses to periodic drives. These responses are robust against small driving imperfections, and are eternal in the thermodynamic limit.
In addition to the external drive, one can induce non-equilibrium dynamics through quantum quench to form a DTC in a disorder-free system with Hermitian~\cite{huang2018clean,yousefjani2025NHDTC} or even non-Hermitian physics~\cite{yousefjani2025NHDTC}.
In closed systems, the existence of DTC order hinges on mechanisms that prevent energy absorption from external driving pulses, such as many-body localization~\cite{khemani2016phase,else2016floquet,yao2017discrete,ho2017critical,khemani2019brief,else2020discrete,Iemini2023}, 
bosonic self-trapping \cite{Wuster2012,Sacha2015,russomanno2017floquet,Giergiel2018c,Matus2019,pizzi2021higher,SachaTC2020}, 
Stark gradient fields \cite{liu2023discrete,kshetrimayum2020stark}, gradient interaction \cite{yousefjani2025DTC}, 
domain-wall confinement~\cite{Collura2022Discrete}, 
and quantum scars \cite{Maskara2021Discrete,Deng2023Using,Bull2022Tuning,Huang2022Discrete,Huang2023Analytical,Bao2024}.

A well-known characteristic of DTCs is the formation of paired GHZ-type states with a $\pi$ quasienergy gap in their spectrum. The emergence of these large-scale entangled states that are responsible for subharmonic responses makes DTCs a promising candidate for quantum technological tasks.  
Thus far, DTCs have found applications in simulating complex systems~\cite{estarellas2020simulating}, topologically protected quantum computation~\cite{bomantara2018simulation}, quantum engine design~\cite{carollo2020nonequilibrium}, and metrology~\cite{lyu2020eternal,Iemini2023,yousefjani2025DTC,moon2024discrete,tsypilnikov2025exact}. 
Sensing in fully connected graphs~\cite{lyu2020eternal} and gradient interacting systems~\cite{yousefjani2025DTC} 
demonstrates the potential of DTCs for quantum-enhanced precision by achieving $\beta{=}3$ in sensing the exchange coupling. While this surpasses the standard quantum limit, the ultimate limit determined by $||H_1||^2{\propto} L^\beta$ with $\beta{=}4$ 
in sensing the strength of two-body interactions, has not been obtained.    
Relying on the long coherence time of DTCs, theoretical~\cite{Iemini2023} and experimental~\cite{moon2024discrete} results demonstrate the capability of DTCs for constructing highly frequency-selective sensors of AC magnetic fields.
Nevertheless, effectively exploiting the inherent GHZ-type eigenstates in the spectrum of the DTCs is a challenge that has yet to be addressed. 
% {\color{red} I TRIED TO REVISE THIS PARAGRAPH BUT IT IS BADLY WRITTEN. HEISENBERG LIMIT IS NOT DEFINED! THESE LIMITS (STANDARD AND HEISENBERG) ARE NOT SPECIAL TO DTCS AND THUS SHOULD BE DEFINED EARLIER.}
% {\color{red} ONE PARAGRAPH FOR periodic-field SENSING IS MISSING. IT HAS ITS OWN LITERATURE.} 

In this work, through concrete theoretical and numerical analysis, we show that a devised probe in the DTC order can achieve ultimate estimation precision in a gradient periodic magnetic field sensing. 
The external periodic-field breaks site invariance symmetry and eliminates the DTC order at a rate that depends on the probe size.
However, for small values of the periodic-field, our results demonstrate that in the DTC order, the probe can benefit from the inherent GHZ state in the spectrum and achieve ultimate precision. 

The robustness of this performance has been analyzed against various types of imperfections and dephasing noise. 
Interestingly, we find that in the sensing scheme, a simple configuration measurement can follow the scaling behavior of the QFI, concerning both time and
size. 
We further provide details of our probe's experimental implementation on ultra-cold atoms in optical lattices.

\section{Quantum parameter estimation }
In this section, we review the theory of quantum parameter estimation. 
The goal is to estimate an an unknown parameter, $h_a$, encoded in the quantum state $\rho(h_a)$.  
The key objective is to minimize the uncertainty in estimating $h_a$, represented by the standard deviation $\delta h_a$. 
This uncertainty is fundamentally bounded by the Cram\'{e}r-Rao inequality~\cite{nla.cat-vn81100, Rao1992}, which relates the minimum error attainable by any locally unbiased estimator to the
classical Fisher information (CFI) 
$\mathcal{F}_C( h_a)$, as 
\begin{equation}
 \delta h_a{\geq}\frac{1}{\sqrt{M\mathcal{F}_{C}( h_a)}}.   
\end{equation}
where $M$ represents the number of repetitions of the sensing protocol.  
For a given set of Positive Operator-Valued Measurement (POVM) operators $\{\Pi_r\}$, the probability of a measurement outcome $r$ is given by $p_r(h_a){=}{\rm Tr} [\rho(h_a)\Pi_r]$. One can obtain the CFI as $\mathcal{F}_{C}(h_a){=}\sum_{r}p_r(h_a) [\partial_{h_a}\ln p_r( h_a)]^2$, with $\partial_{h_a}$ being the derivatives with respect to the parameter $h_a$.
Optimizing over all possible POVMs $\{\Pi_r\}$ leads to the quantum Cram\'{e}r-Rao inequality, which is bounded by the QFI $\mathcal{F}_Q( h_a)$
as 
\begin{equation}\label{Eq.QCR}
\delta h_a{\geq}\frac{1}{\sqrt{M\mathcal{F}_{C}( h_a)}} {\geq}\frac{1}{\sqrt{M\mathcal{F}_{Q}( h_a)}}.  
\end{equation}
Thus, the QFI, as a central quantity in metrology, determines the ultimate precision limit in quantum sensing.
For pure quantum states $\rho(h_a) {=} |\psi( h_a)\rangle\langle\psi( h_a)|$, the QFI takes a simple form as~\cite{fisher1922mathematical}
\begin{equation}\label{Eq.QFI}
\mathcal{F}_{Q}( h_a){=}4 \left(\langle \partial_{ h_a}\psi( h_a) | \partial_{ h_a}\psi( h_a) \rangle{-}|\langle \partial_{ h_a}\psi( h_a) | \psi( h_a) \rangle|^2 \right).    
\end{equation}
The performance of a sensor is characterized by how the QFI scales with available resources, such as probe size and time. 
In particular, in closed systems, this relationship can be expressed as $\mathcal{F}_Q {\sim}n^{\alpha} L^\beta$, where $n$ and $L$
represent the duration time of the protocol and the size of the probe, respectively, with $\alpha$ and $\beta$ as the corresponding exponents.
While, without quantum effects, at best one can achieve $\alpha{=}\beta{=}1$, exploiting quantum features, such as entanglement and squeezing, enhances the scaling to $\alpha{>}1$ and $\beta{>}1$. 
A notable case is when $\alpha{=}2$ and $\beta{=}2$, which is called Heisenberg scaling. 
The most common way to capture this scaling is by initializing a dynamical protocol in maximally entangled states, such as Greenberger–Horne–Zeilinger (GHZ) states, and encoding the desired parameter through a unitary evolution, while the interaction between the system's components remains negligible~\cite{giovannetti2004quantum, leibfried2004toward,giovannetti2006quantum,banaszek2009quantum,giovannetti2011advances,frowis2011stable,demkowicz2012elusive,wang2018entanglement,kwon2019nonclassicality,toth2014quantum}.
%The limitations of this method prompt efforts to unlock the potential of many-body interacting systems.
Alternatively, strongly correlated many-body systems can also be used a s a resource for achieving quantum enhanced sensitivity  by harnessing a variety of quantum features such as criticality~\cite{montenegro2024Rev.}. In this type quantum sensors, unlike GHZ-type sensors, the interaction between the components of the probe plays a crucial role in obtaining quantum enhanced sensitivity. Nonetheless, the obtainable precision is fundamentally bounded by the interaction details in the system~\cite{boixo2007generalized,abiuso2025fundamental,puig2024dynamical}. 

\section{Model }
\begin{figure}
    \centering
\includegraphics[width=0.45\linewidth,height=0.36\linewidth]{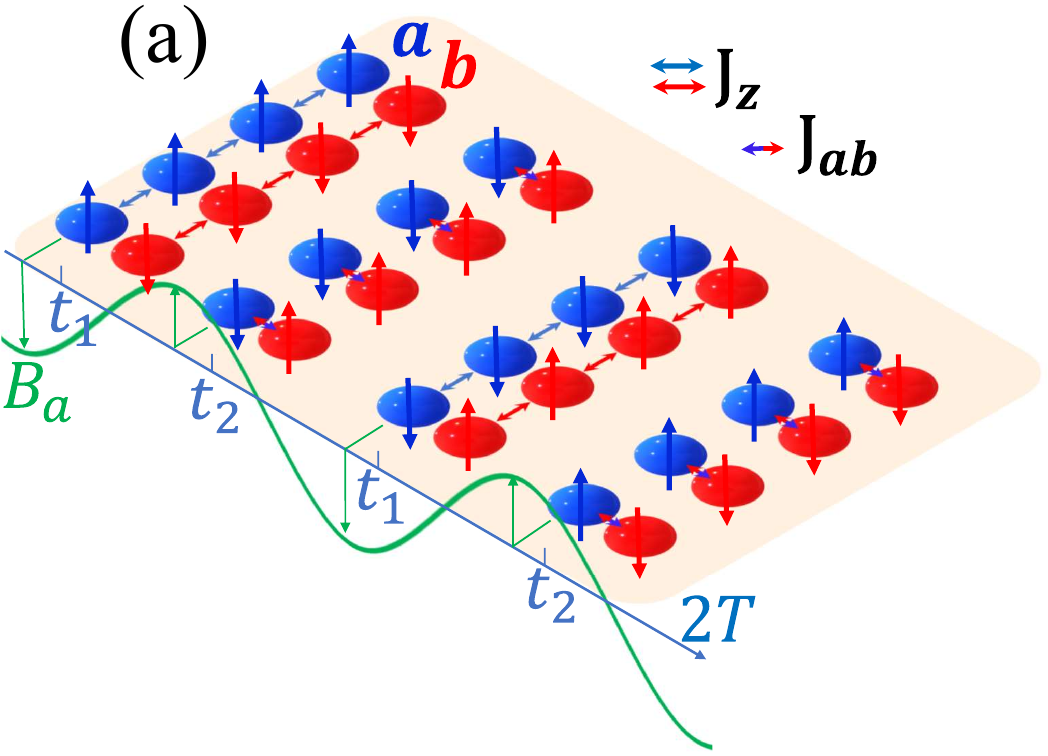}
\includegraphics[width=0.54\linewidth,,height=0.45\linewidth]{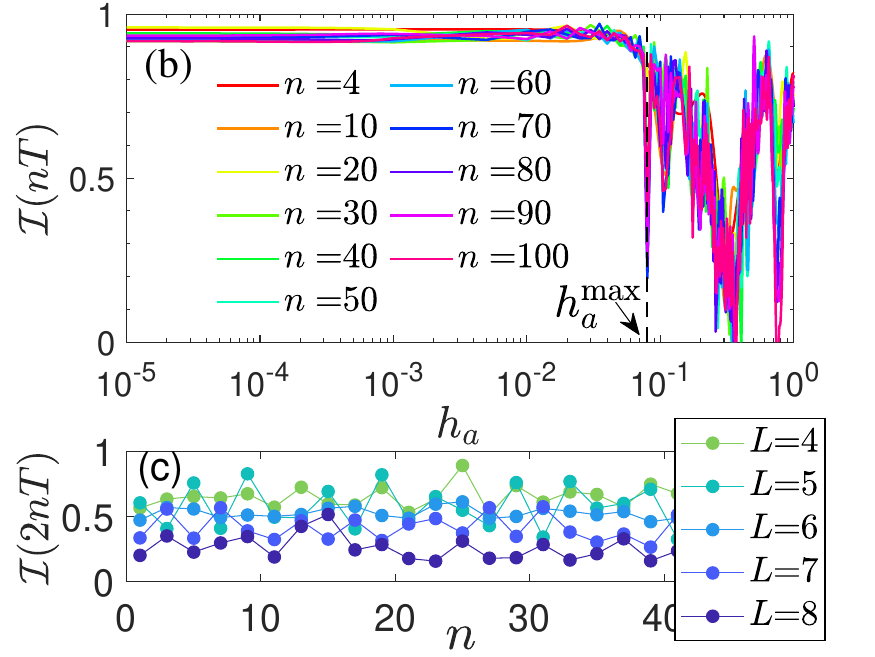}
    \caption{(a) Schematic for the DTC sensor introduced here. (b) Imbalance in the total magnetization of the chains $a$ and $b$ denoted as $\mathcal{I}(nT)$, versus the amplitude of the periodic-field $h_{a}$ at different stroboscopic times in a system of size $L{=}6$. (c) Imbalance at even stroboscopic times in systems with various sizes, when $h_a{=}0.25$ and the system operates in the non-DTC region. In both panels, we set $\varepsilon{=}0.1$. }
    \label{fig:Fig1}
\end{figure}
Our system consists of two one-dimensional chains, labeled $a$ and $b$, each containing $L$ spin-$1/2$ particles. The system is under binary quench and the Hamiltonian reads 
\begin{equation}\label{Eq.Hamiltonian}
H(t) = 
\begin{cases}
H_a+H_b=-J_z\sum_{\mu=a}^b\sum_{j=1}^{L-1}\sigma_j^{\mu z}\sigma_{j+1}^{\mu z} & 0<t\leqslant t_1 \\
H_{I} = J_{ab}\sum_{j=1}^{L} ( \sigma_j^{a+}\sigma_{j}^{b-} +  \sigma_j^{a-}\sigma_{j}^{b+}) & t_1<t\leqslant t_1 + t_2
\end{cases}  
\end{equation}
where $J_z$ denotes the spin exchange coupling, $\sigma^{\mu\pm}_{j}{=}\frac{1}{2}(\sigma^{\mu x}_{j} {\pm} i\sigma^{\mu y}_{j})$, and $\sigma^{\mu(x,y,z)}_{j}$ representing
Pauli operators acting at site $j$ of chain $\mu{=}a,b$.
We set $J_{ab}{=}\pi J_z(1{-}\varepsilon)$, with $\varepsilon{\in}\mathds{R}$ as a small deviation from a perfect coupling between two chains and $t_1{=}t_2{=}1{/}2J_z$. 
This binary quench creates a time-crystalline temporal order with periodicity $T_{dtc}{=}2T$ in which $T{=}t_1{+}t_2$, in a system initialized in $|\psi(0)\rangle{=}|{\uparrow}{\cdots}{\uparrow}\rangle_{a}{\otimes}|{\downarrow}{\cdots}{\downarrow}\rangle_{b}$ ~\cite{huang2018clean,yousefjani2025NHDTC}.
This means that, for infinitesimal $\varepsilon$, 
% the population of two different energy levels of $H_a{+}H_b$ switches at half drive frequency, a phenomenon known as subharmonic locking. 
measuring the imbalance in the total magnetization of chains $a$ and $b$, 
\begin{equation}\label{Eq:IM}
  \mathcal{I}(nT) = \frac{1}{\mathcal{I}(0)} \sum_{j=1}^{L}\langle \psi(nT)|(\sigma_{j}^{az}{-}\sigma_{j}^{bz})|\psi(nT)\rangle  
\end{equation}
shows persistent coherent oscillation between 
$\mathcal{I}(nT){=}1$ at even $n$'s obtained for $|\psi(0)\rangle$, 
and $\mathcal{I}(nT){=}{-}1$ at odd $n$'s obtained for the inversion partner of the initial state $|\tilde{\psi}(0)\rangle{=}\Pi_{j=1}^{L}\sigma^{ax}_{j}{\otimes}\sigma^{bx}_{j}|\psi(0)\rangle$. 
The origin of this period-doubling oscillations that last for $\tau{\propto} e^{\alpha L}$ with $\alpha{=}\ln(\varepsilon^{-1.033})$ is the emergence of $\pi$-Paired Floquet states as $|\Phi_{\pm}\rangle{\cong}(|\psi(0)\rangle{\pm}|\tilde{\psi}(0)\rangle){/}\sqrt{2}$ 
with eigenvalues differing by $\pi$~\cite{yousefjani2025NHDTC}.
 The sensing capability of general DTCs for estimating a uniform periodic-field has been investigated in Ref.~\cite{tsypilnikov2025exact}. The analytical calculations reveal the potential of capturing the quantum enhancement with respect to the system size as $\propto L^2$. 
One can expect to obtain the same results in sensing a uniform magnetic field using the presented system in Eq.~\ref{Eq.Hamiltonian}. This motivates us to investigate the capability for sensing a non-uniform field. To this end, we expose the chain $a$ to a gradient periodic-field in the $z$-direction as
\begin{equation}\label{Eq.H_AC}
 H_{ac}(t)=
 B_a(t)G^{az} = h_{a}\sin(2\pi f_{ac}t)\sum_{j=1}^{L}j\sigma_{j}^{a z}, 
\end{equation}
where $h_a$ and $f_{ac}$ are the amplitude and frequency of the periodic-field, respectively. 
See Fig.~\ref{fig:Fig1}(a) for a system schematic. 
%The cross-talk effect will be discussed later. 
To trace the effect of this periodic-field on dynamics, we approximated $B_a(t)$ with a square-pulse form of the same period and alternating sign. This approximation is justified in our case because our focus is on the accumulated phase over time rather than the detailed structure of the continuous driving field. This approximation captures the essential contribution of the periodic-field to the dynamics.
Under this assumption, $H_{\text{ac}}(t)$ becomes a piecewise static Hamiltonian that flips sign every period $T_{ac}$. For a periodic-field that resonates with the DTC, namely $T_{ac}{=}2T$, the dynamics can be well described by an effective Hamiltonian as 
\begin{equation}\label{Eq.Eff_Ham}
H_{\mathrm{eff}}(t) = 
\begin{cases}
H_a+H_b+\Theta_{n}G^{az} & (n{-}1)T<t\leqslant (n{-}\frac{1}{2})T \\
H_{I} + \Theta_{n}G^{az} & (n{-}\frac{1}{2})T<t\leqslant nT
\end{cases}  
\end{equation}
with $   \Theta_n {=} h_a{\int_{(n-1)T}^{(n-\frac{1}{2})T}\sin(\frac{\pi \tau}{T})d \tau} {=}
   h_a\int_{(n-\frac{1}{2})T}^{nT}\sin(\frac{\pi\tau}{T})d \tau  {=}(-1)^{n+1}\frac{h_{a}}{\pi J_z}$, 
% \begin{align}\label{Eq.theta}
%    \Theta_n {=} h_a{\int_{(n-1)T}^{(n-\frac{1}{2})T}\sin(\frac{\pi \tau}{T})d \tau} {=}
%    h_a\int_{(n-\frac{1}{2})T}^{nT}\sin(\frac{\pi\tau}{T})d \tau  {=}(-1)^{n+1}\frac{h_{a}}{\pi J_z},   
% \end{align}
as the time-averaged strength of the periodic-field.
This allows us to express the Floquet operator as 
\begin{align}\label{Eq.Propagator2}
U_n {\simeq} 
e^{-i(H_{I} + \Theta_{n} G^{az})} e^{-i((H_a + H_b)+\Theta_{n} G^{az})}.
\end{align}
Here we set $J_z{=}1$ as the unit of energy.
During the first period, $t \in [0, T]$, the spins in the chain $a$ are primarily in $|{\uparrow,\cdots,\uparrow}\rangle_a$, and the field is positive. The accumulated phase during this interval is therefore positive.
During the second period, $t \in [T, 2T]$, the field becomes negative, but the spins flip to $|{\downarrow,\cdots,\downarrow}\rangle_a$, leading again to a positive phase accumulation.
To show the effect of the implemented periodic-field on the DTC order, Fig.~\ref{fig:Fig1}(b) plots the imbalance $\mathcal{I}(nT)$ versus the amplitude of the periodic-field $h_{a}$ over tens of stroboscopic times $n$ when the system initialized in $|\psi(0)\rangle$.
 By increasing $h_{a}$, the perfect and stable oscillations of the imbalance in stroboscopic times, namely  $\mathcal{I}(2nT){=}1$, 
are replaced by nontrivial oscillations.
This hints that our system goes through a sharp phase transition from a stable DTC order (for $h_a{<}h_a^{\max}$) to a region with no spontaneous breaking of DTTS  (for $h_a{>}h_a^{\max}$), denoted as non-DTC region.
Surprisingly, this transition,  happening in a specific value denoted by $h_{a}{=}h_{a}^{\max}$ (dashed line), repeats in all the stroboscopic times, and, as we show later, is robust against the deviation parameter $\varepsilon$.
In Fig.~\ref{fig:Fig1}(c), we plot the dynamic of $\mathcal{I}(2nT)$ when $h_a{=}0.25$ and hence, the systems of different sizes operate in the non-DTC region.
Clearly, by enlarging the system size, the amplitude of the  
nontrivial oscillations decreases considerably. This implies that, in systems of large enough sizes, these oscillations practically vanish even in the early
time windows, signaling the thermalization of the system.

In the following, we first show that the DTC phase can serve as a valuable resource for quantum-enhanced sensitivity and provide ultraprecise periodic-field sensing. Then, we complete our study by providing a robustness analysis of this enhancement to the perturbations in initialization, frequency offset, and crosstalk. We also explore the effect of $\varepsilon$ on the capability of our probe and propose an experimental protocol for its implementation.    

\section{DTC sensor } 
\begin{figure}
    \centering
    \includegraphics[width=0.49\linewidth]{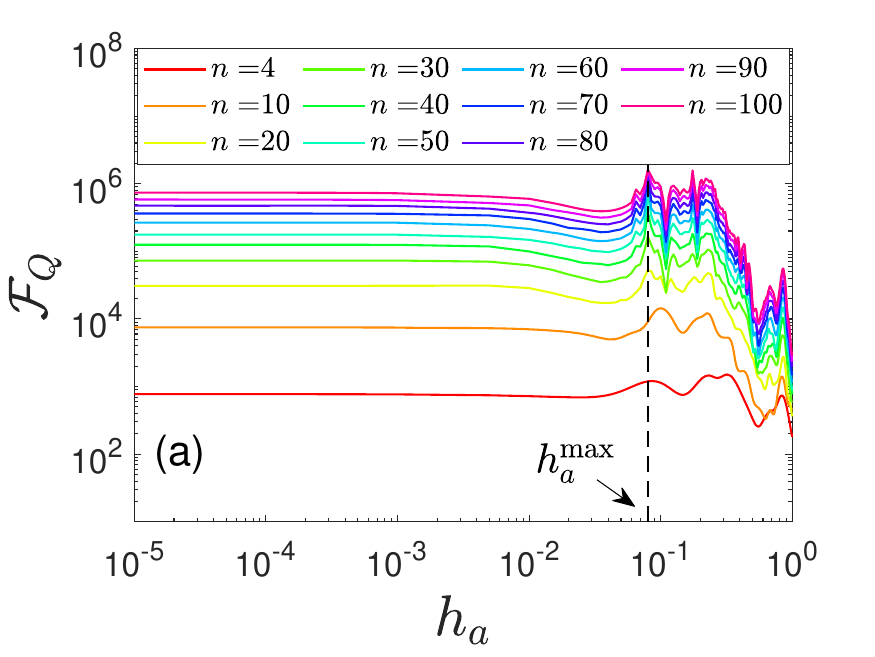} 
    \includegraphics[width=0.49\linewidth]{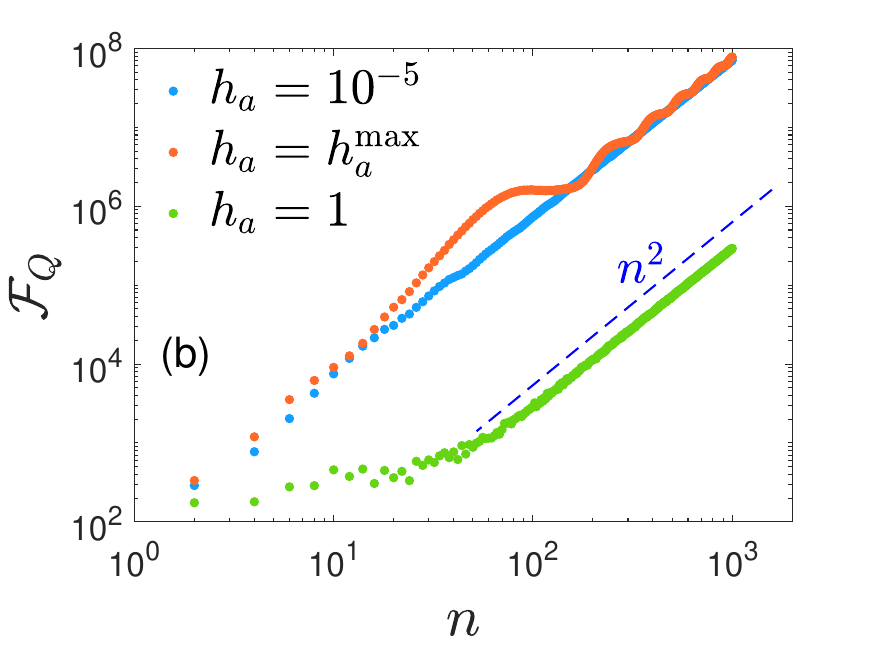}
    \includegraphics[width=0.49\linewidth]{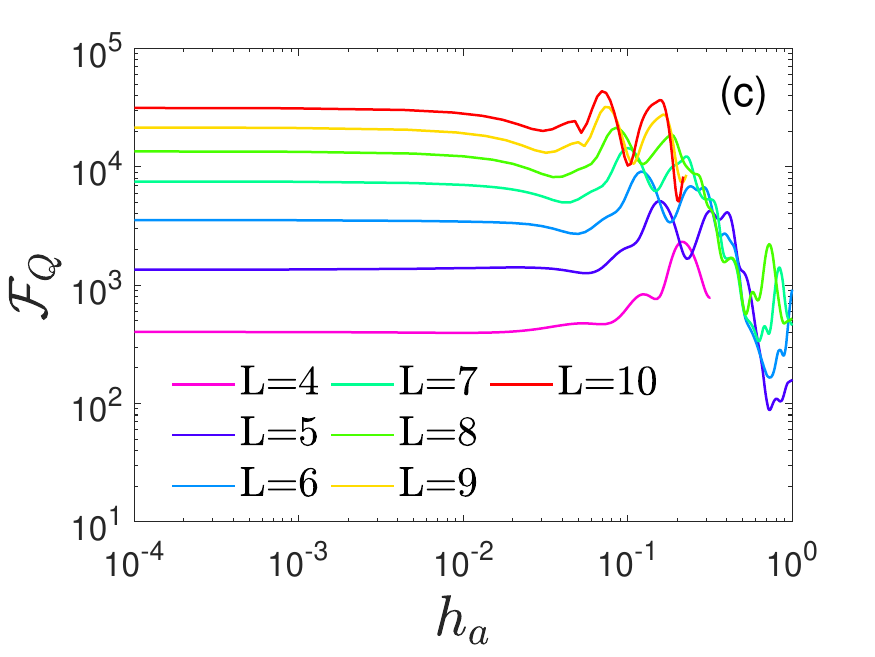}
    \includegraphics[width=0.49\linewidth]{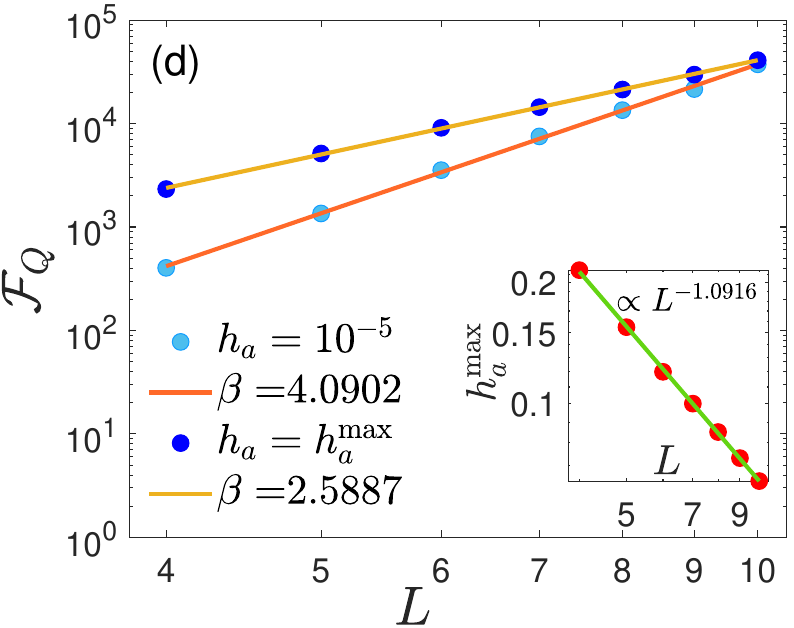}
    \caption{(a) The QFI as a function of $h_a$ over different stroboscopic times. (b) The dynamical behavior of the QFI versus stroboscopic time $n$ at different values of $h_{a}$. In both panels (a) and (b), we set $L{=}7$ and $\varepsilon{=}0.1$. (c) QFI as a function of $h_a$ for systems of different sizes over $n{=}10$ period cycles when $\varepsilon{=}0.1$. (d) The values of QFI inside the DTC phase, namely $h_{a}{=}10^{-5}$, and in the transition point, namely $h_a{=}h_{a}^{\max}$, as a function of the system size $L$. The markers are the numerical results, and the solid lines provide the best fitting function as $\mathcal{F}_{Q}{\propto}L^{\beta}$ with $\beta{=}4.0902$ and $\beta{=}2.5887$ for a sensor in the DTC phase and in its transition point. The inset is the transition point $h_{a}^{\max}$ as a function of $L$. The best fitting function is obtained as $h_{a}^{\max}{\propto}L^{-1.0916}$.}
    \label{fig:Fig2}
\end{figure}
To assess the performance of our DTC sensor for measuring the amplitude of the periodic-field, one can show that $U_t{=}\mathcal{T}e^{-i\int_{0}^{nT}(H(t)+H_{ac}(t))dt}
{=}\Pi_{k=1}^{n}U_k$ for $t{=}nT$ while $U_{k}$ is the propagator for one period cycle given in Eq.~(\ref{Eq.Propagator2}).
 
For a resonant periodic-field with the DTC, straightforward calculations lead to 
\begin{align}\label{Eq.derivative of U}
 \partial_{h_{a}} U_{t} &= {-i}U_{t} \sum_{k=1}^{n}  \Big(\Pi_{l=1}^{k} U_l \Big)^{\dagger} \Big(\partial_{h_{a}}\Theta_{k}G^{az}\Big) \Big(\Pi_{l=1}^{k} U_l \Big) \leqslant i U_t n G^{az}/\pi.
\end{align}
Here, we used the following simplification 
\begin{align}\label{Eq.Simplification}
  -i \Big(\partial_{h_{a}}&\Theta_{k}\Big) \Big(\Pi_{l=1}^{k} U_l \Big)^{\dagger} G^{az} \Big(\Pi_{l=1}^{k} U_l \Big) \cr &=
   \begin{cases}
      i(-1)^{k+1} G^{az}/\pi & (k-1)T<t_{k} \leqslant (k-1/2)T \\
       i(-1)^{k+1}D_k^{az}/\pi & (k-1/2)T< t_{k} \leqslant kT \\
   \end{cases}
\end{align}
wherein $D_k^{az}{=}\sum_{j=1}^{L} ja_j^k\sigma^{az}_j{\leqslant}\sum_{j=1}^{L} j\sigma^{az}_j{=}G^{az}$ and
\begin{equation}\label{Eq.alpha}
 a_j^{k}{=}\frac{j^2 \Theta_k^2+J_{ab}^2\cos^2\big(\sqrt{j^2 \Theta_k^2 + J_{ab}^2}\big)}{j^2 \Theta_k^2+J_{ab}^2}.   
\end{equation}
Using
\begin{align}
 \mathcal{F}_{Q}(h_a)&= \\ 
 4&\left(\langle \psi(0)| (\partial_{h_{a}} U_t)^{\dagger}(\partial_{h_{a}} U_t)|\psi(0) \rangle{-}|\langle \psi(0)|  U_t^{\dagger}(\partial_{h_{a}} U_t)|\psi(0) \rangle|^2 \right) \nonumber
\end{align}
and 
 Eq.~\ref{Eq.derivative of U}, one can simply obtain $\mathcal{F}_{Q}(t) {\leqslant} 4 n^2(\Delta G^{az})^2/{\pi^2} $ in which $\Delta G^{az}{=}\sqrt{\langle (G^{az})^2 \rangle {-} \langle G^{az} \rangle^2}$ stands for the variance.
 This bound depends on the variance of $G^{az}$ in the initial state. If the initial state has vanishing variance with respect to this generator, i.e. $\Delta G^{az} = 0$, then the QFI bound becomes zero. This outcome is consistent with the fact that, in such a case, the state is an eigenstate of $G^{az}$. As a consequence, the dynamics generated by $G^{az}$ cannot imprint any distinguishable phase information on the state, and thus no sensitivity can be gained, leading to a vanishing QFI throughout the evolution.
 However, one can use the upper bound of the variance, defined by the square of the spectral norm defined as $||G^{az}||^2{=}(\lambda_{\max}{-}\lambda_{\min})^2/4$ with $\lambda_{\max}{=}L(L{+}1)/2$ and $\lambda_{\min}{=}{-}L(L{+}1)/2$ as the largest and smallest eigenvalues of $G^{az}$. 
This results in $\mathcal{F}_{Q}{\leqslant}n^2L^2 (L+1)^2/\pi^2.$ 
 Capturing this upper bound that determines the ultimate precision in estimating $h_{a}$ requires satisfying two conditions. 
First, $a_j^k{\simeq}1$. Concerning Eq.~\ref{Eq.alpha}, one see that $\varepsilon{=}0$ and $h_a{\ll} \sqrt{2}\pi^2/L$ can lead to $\cos^2\Big(\sqrt{j^2 \Theta_k^2 + J_{ab}^2}\Big){\simeq}1$.
Second, the probe is to be initialized in the balanced superposition of the eigenstates corresponding to the extremal eigenvalues of $G^{az}$, namely \(|\lambda_{\max}\rangle{=}|{\uparrow,\cdots,\uparrow}\rangle_a\) and 
$|\lambda_{\min}\rangle{=}|{\downarrow,\cdots,\downarrow}\rangle_a$.
Therefor, the optimal initial state is $\frac{1}{\sqrt{2}}\big(|{\uparrow,\cdots,\uparrow}\rangle_a{\otimes} |{\downarrow,\cdots,\downarrow}\rangle_b {\pm}|{\downarrow,\cdots,\downarrow}\rangle_a{\otimes} |{\uparrow,\cdots,\uparrow}\rangle_b\big)$, concerning the conservation of the total magnetization in our protocol.
In the following, we show that, although initialized in the product state $|\psi(0)\rangle$,  achieving this ultimate precision even for $\varepsilon{\neq}0$ is permitted in our DTC probe, due to the freely emerged "Schr{\"o}dinger cat" state $(|\psi(0)\rangle{\pm}|\tilde{\psi}(0)\rangle){/}\sqrt{2}$ in the spectrum of DTC.

Fig.~\ref{fig:Fig2}(a) presents the QFI as a function of $h_a$ for various cycling periods when the periodic-field is in resonance with the DTC, namely $T_{ac}{=}T$. 
Obviously, the QFI behaves differently in each phase. While in the DTC phase, one observes a plateau that uniformly increases with $n$, in the non-DTC region, non-trivial oscillations and suppression of the QFI are evident. 
The QFI clearly peaks at the same transition point denoted by $h_{a}^{\max}$ (dashed line) in Fig.~\ref{fig:Fig1}(b). The peaks become sharper as $n$ increases. 
Fig.~\ref{fig:Fig2}(b) depicts the dynamics of the QFI in three various regions, deep inside the DTC ($h_a{=}10^{-5}$), at the transition point ($h_a{=}h_{a}^{\max}$) and deep inside the non-DTC ($h_a{=}1$). 
The results show the square growth of the QFI, namely $\mathcal{F}_{Q}{\propto}n^2$ in the DTC phase. The unlimited quantum correlations at the transition point result in a faster growth of the QFI over early multiple cycles. However, at later times, one has $\mathcal{F}_{Q}{\propto}n^2$. In the non-DTC region, the origin of QFI growth in later time is the contribution of several Floquet states that have considerable overlap with $|\psi(0)\rangle$~\cite{yousefjani2025NHDTC}.
To extract the scaling behavior of the QFI, in Fig.~\ref{fig:Fig2}(c) we plot the obtained QFI as a function of $h_a$ after $n{=}10$ cycling period when the size of the chains varies. 
In both DTC and the transition point, the finite-size effect is apparent. By enlarging the chains, the peak of the QFI as the indicator of the phase transition skew toward smaller values of $h_{a}$ with a rate of $h_{a}^{\max}{\propto}L^{-1.0916}$, as shown in Fig.~\ref{fig:Fig2}(d). 
The result aligns with the analytical requirements of $h_a{\ll}\sqrt{2}\pi^2{/}L$ to achieve the ultimate precision.
 This means that, in the thermodynamic limit ($L \to \infty$), even an infinitesimal periodic field can destabilize the DTC order. While this implies fragility of the DTC phase in large systems due to the presence of the periodic-field, it simultaneously enables the system to act as an extremely sensitive detector, as vanishingly small field produces a pronounced dynamical response, effectively amplifying the system’s susceptibility to weak signals.
This sharp transition in the dynamics can be exploited to detect the presence and properties of extremely weak periodic fields, making the system well-suited for high-precision metrology in practice.
% Although this result indicates degradation of the DTC order in the thermodynamic limit due to the presence of the periodic-field, in practice, it facilitates ultraprecise sensing of infinitesimal periodic-fields.
In Fig.~\ref{fig:Fig2}(d), we show the values of the QFI deep inside the DTC phase ($h_{a}{=}10^{-5}$) and at the transition point ($h_a{=}h_{a}^{\max}$) when the length of the chains varies. The numerical results (markers) are well described by fitting the function $\mathcal{F}_{Q}{\propto}L^{\beta}$ with $\beta{=}4.0902$ and $\beta{=}2.5887$ for a sensor working inside the DTC order and at the transition point, respectively. 
This evidence indicates that leveraging a DTC system that is initialized in a product state $|\psi(0)\rangle$ results in a significant quantum enhancement in the sensing of the periodic-field, 
represented by 
\begin{equation}
    \mathcal{F}_{Q}\propto L^{\beta}n^\alpha \quad \mathrm{with} \quad \beta{\simeq}4, \quad \alpha{\simeq}2.
\end{equation}

To understand the origin of this quantum enhancement, one can focus on the energy spectrum of the period-doubling DTCs, namely the emergence of the "Schr{\"o}dinger cat" states~\cite{khemani2016phase,russomanno2017floquet,estarellas2020simulating,yousefjani2025NHDTC}. 
In the absence of the periodic-field, the spectrum $U_n$ in our DTC has a pair of eigenstates with strong overlap with $|\psi\rangle$ and $|\tilde{\psi}\rangle$~\cite{yousefjani2025NHDTC}. 
These eigenstates can be well approximated by the long-range correlated "Schr{\"o}dinger cat" states $|\Phi_{\pm}\rangle{\cong}\frac{1}{\sqrt{2}}(|\psi(0)\rangle{\pm}|\tilde{\psi}(0)\rangle)$, and the $\pi$ quasienergy gap between them causing the subharmonic response in dynamic form $|\psi(0)\rangle$ or $|\tilde{\psi}(0)\rangle$. For a small magnetic field, the spectrum of $U_n$ remains unchanged, and the sensing protocol can benefit from the freely generated maximally entangled states. 

\section{Robustness analysis }
Having elucidated the performance of the DTC probe for sensing a periodic-field, we now explore the viability of our probe under various types of defects. 
After relaxing the criteria of resonance in the periodic-field, we follow up by studying the effect of the crosstalk and perturbed initialization.
We complete the study by analyzing the effect of the imperfect quench. 
\subsection{Off-resonance effect}
\begin{figure}
    \centering
    \includegraphics[width=0.49\linewidth]{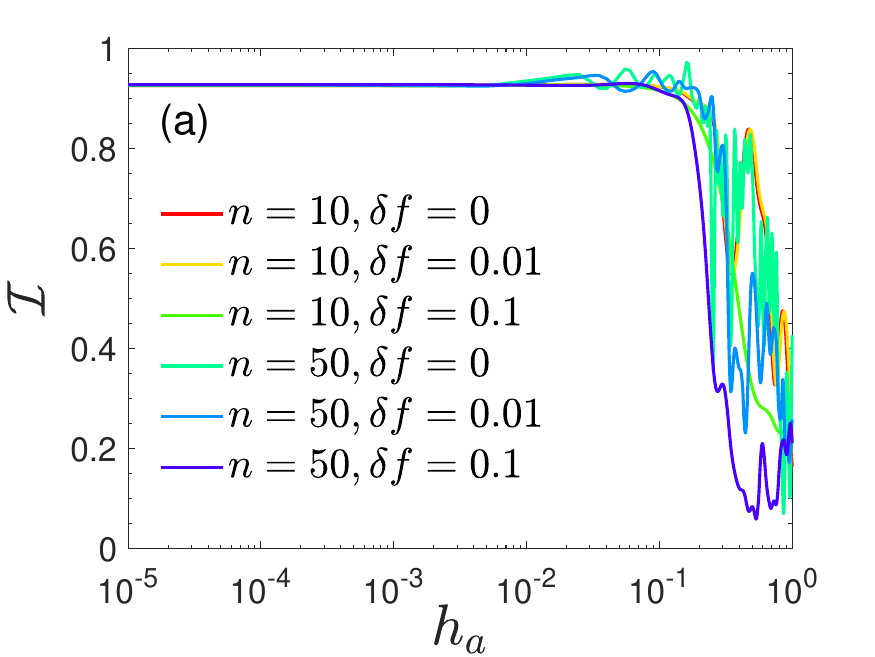} 
    \includegraphics[width=0.49\linewidth]{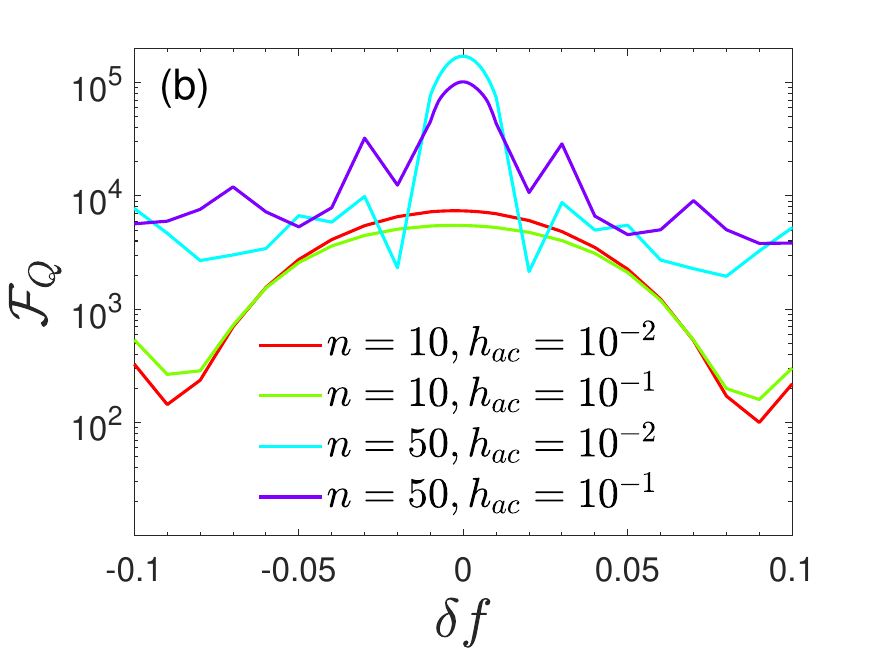} 
    \includegraphics[width=0.49\linewidth]{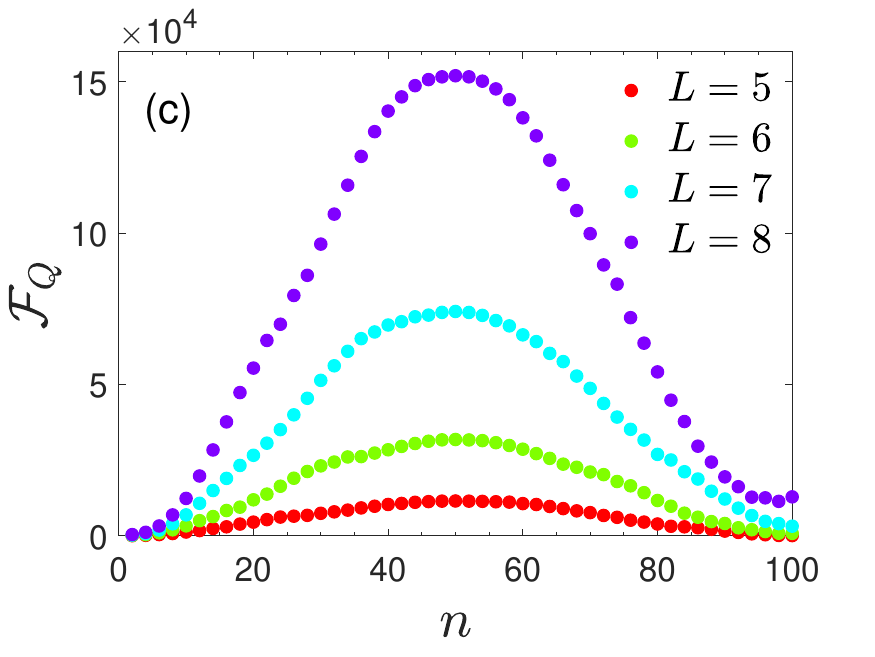} 
    \includegraphics[width=0.49\linewidth]{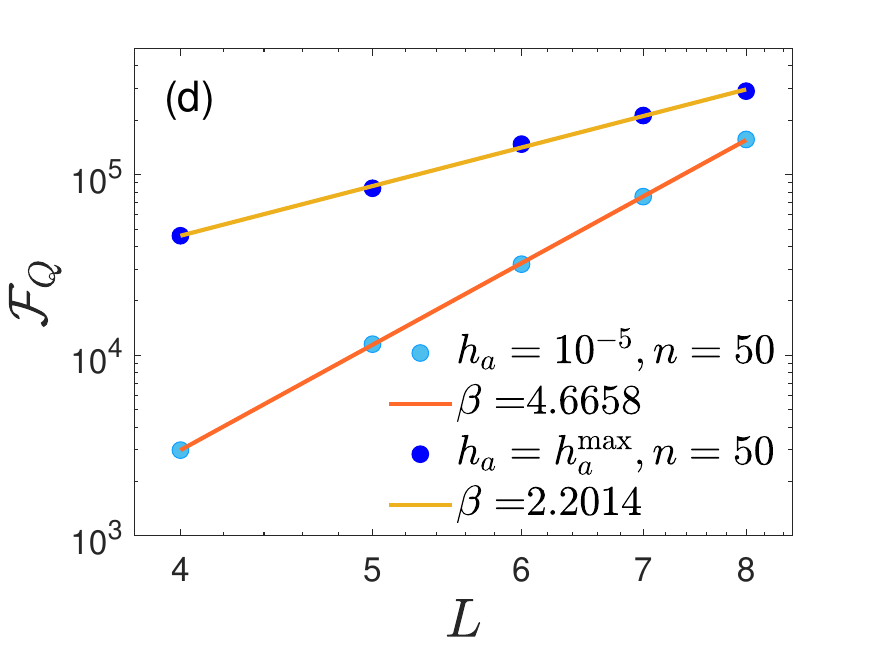} 
    \caption{(a) The imbalance as a function of $h_a$ for different values of the detuning $\delta f$ over different stroboscopic times. (b) The QFI versus the detuning $\delta f$ for various values of the $h_a$ and $n$. In both panels $L{=}7$. (c) The QFI in systems with different sizes as a function of $n$ when $h_{a}{=}10^{-2}$ and $\delta f{=}10^{-2}$. (d) The QFI as a function of $L$ when the sensor is in the DTC phase and at the transition point and $\delta f{=}10^{-2}$. The numerical results are well described by the fitting function $\mathcal{F}_{Q}{\propto}L^\beta$ with $\beta{=}4.6658$ and $\beta{=}2.2014$ for a sensor deep inside the DTC phase and at the transition point, respectively. 
    In all the panels, we set $\varepsilon{=}0.1$.   }
    \label{fig:Fig3}
\end{figure}
 While far off-resonance periodic-field has a negligible impact on the dynamic of the system, slight deviation denoted by $f_{ac}/f{=}1{+}\delta f$ with $\delta f$ as the frequency offset of the DTC's frequency may affect the lifetime of the DTC order and the performance of the probe.
In this case, one can again approximate the sinusoidal periodic-field with a square-wave form of the same period and obtain the total accumulated phase through integrating over $h_a \sin(2\pi (1+\delta f)f \tau)$ over the two halves of each period $T$, namely $[{(n-1)T},{(n-\frac{1}{2})T} ]$ and $[{(n-\frac{1}{2})T},{nT}]$.
Therefor, the total accumulated phase in each period cycle $T$ is $\frac{(-1)^{n+1} h_a}{\pi(1+\delta f)}\big(\cos(n\pi\delta f){+}\cos((n{-}1)\pi\delta f)\big)$.
In Fig.~\ref{fig:Fig3}(a) we present the imbalance as a function of $h_a$
over multiple period cycles when $\delta f$ varies. 
The response of $\mathcal{I}(nT)$ in the DTC region reveals the robustness of the system to small offset values. 
Fig.~\ref{fig:Fig3}(b) plots the QFI as a function of $\delta f$ over multiple period cycles when the probe works inside the DTC order.
Clearly, the maximum values for the QFI happen for a resonant periodic-field.
By increasing $\delta f$, the unfavorable off-resonance correlations affect the performance of the probe.
Note that $\Theta_n{\rightarrow}0$ whenever $n\delta f{\rightarrow}0.5$.
This affects the growth of the accumulation phase and therefore the value of the QFI as can be seen in Fig.~\ref{fig:Fig3}(c) where we plot the dynamics of the QFI in systems of various lengths under a periodic-field with magnitude $h_a{=}10^{-2}$ and offset $\delta f{=}10^{-2}$. All the accumulated phase in the first time window, namely for $\delta f n{<}0.5$, will be erased in the second time window, namely for $\delta f n{>}0.5$. 
This introduces an optimal sensing time as $n^{*}{=}1/2\delta f$, giving the peaks of the QFI. Interestingly, sensing in $n{\leqslant}n^{*}$ kept the
scaling behavior of the QFI intact, as evidenced by Fig.~\ref{fig:Fig3}(d), where we plot the obtained $\mathcal{F}_{Q}$ for different system sizes when the probe works inside the DTC and at the transition point with offset $\delta f{=}10^{-2}$. 
Clearly, the obtained scaling $\beta$ is in line with our results for the resonance case, indicating the attainability of the quantum-enhanced sensitivity.  
So far, we have focused on the scenario where the periodic-field is detected exclusively by subsystem $a$.
However, in practice, the close distance between chains $a$ and $b$
introduces crosstalk, causing $b$ also to sense a portion of the periodic-field.
When initializing the subsystems in polarized states, but with opposite total magnetization in the $z$-direction, the crosstalk influences the accumulated phase and, consequently, the value of the QFI. 

\subsection{Cross-talk effect}
\begin{figure}[t!]
    \centering
    \includegraphics[width=0.49\linewidth]{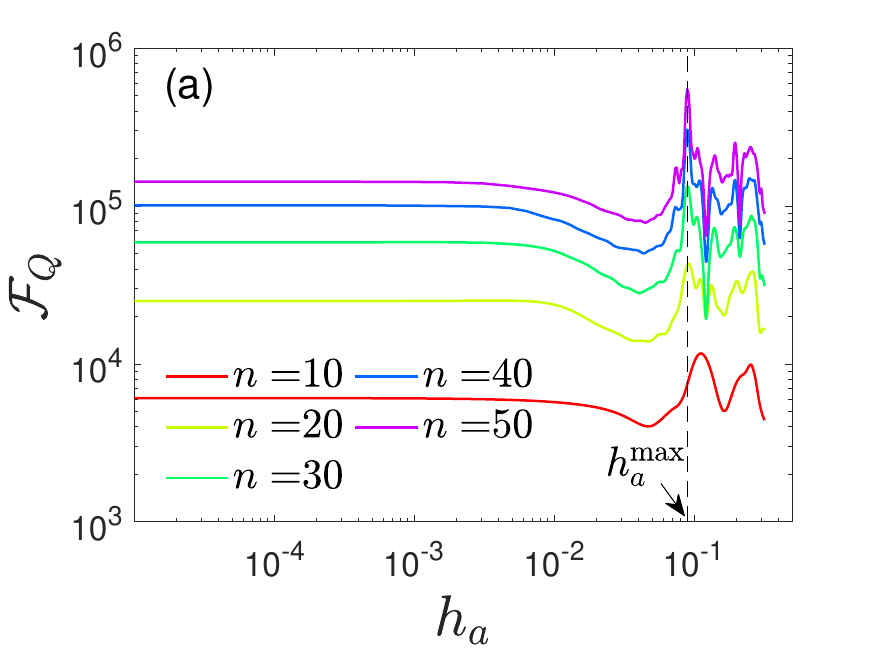}
    \includegraphics[width=0.49\linewidth]{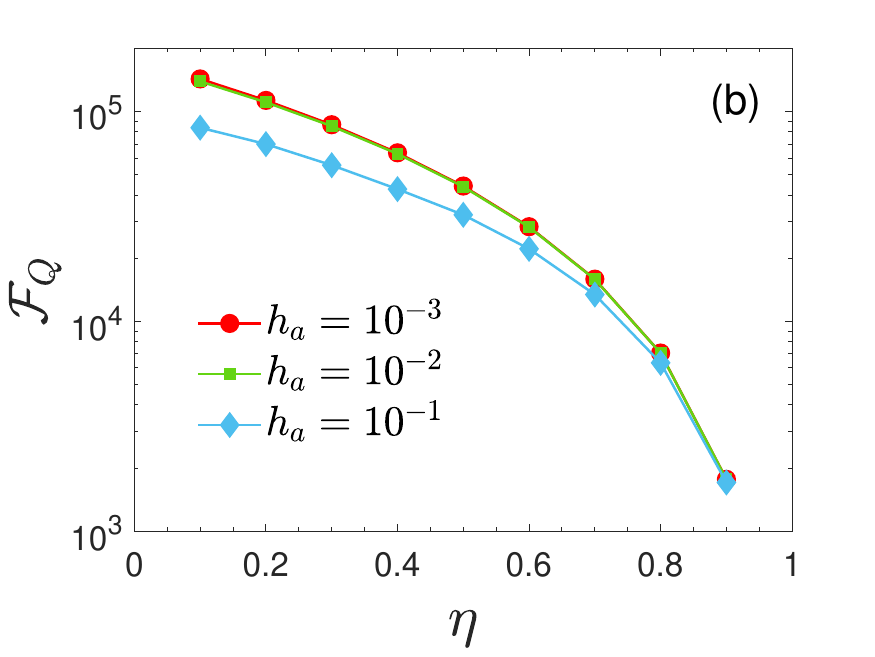}
    \caption{(a) The QFI as a function of $h_a$ over various stereoscopic times, when $\eta{=}10^{-1}$.
    (b) The QFI as a function of $\eta$ when $n{=50}$. In both plots we set $L{=}7$ and $\varepsilon{=}0.1$.}
    \label{fig:CT}
\end{figure} 
To analyze the effect of the crosstalk, we replace the $H_{ac}(t)$ in Eq.~(\ref{Eq.H_AC}) by
\begin{equation}\label{Eq.H_AC_CT}
    H_{ac}(t) = B_a(t)G^{az} + \eta B_a(t)G^{bz}, \quad \mathrm{with}\quad G^{\mu z} =\sum_{j=1}^{L} j \sigma^{\mu z}_{j},
\end{equation}
for $\mu{=}a,b$ in which $\eta{\in}(0,1)$. 
As evidenced by our results in Fig.~\ref{fig:CT}(a), where the QFI, as a function of $h_{a}$, has been plotted at different stroboscopic times when $\eta{=}10^{-1}$, the crosstalk can not affect the DTC order and the phase transition is driven by $h_{a}$.
The distinct behavior of the QFI in both DTC and non-DTC regions, along with the enhanced peaks at the phase transition point, is evident. For completeness, in Fig.~\ref{fig:CT}(b) we present the QFI obtained as a function of $\eta$ when $h_a$ varies. The results are obtained after $n{=}50$ period cycling. 
Clearly, as the crosstalk parameter is increased, the values obtained for QFI decrease. 

\subsection{Initialization effect}
\begin{figure}
    \centering
    \includegraphics[width=0.49\linewidth]{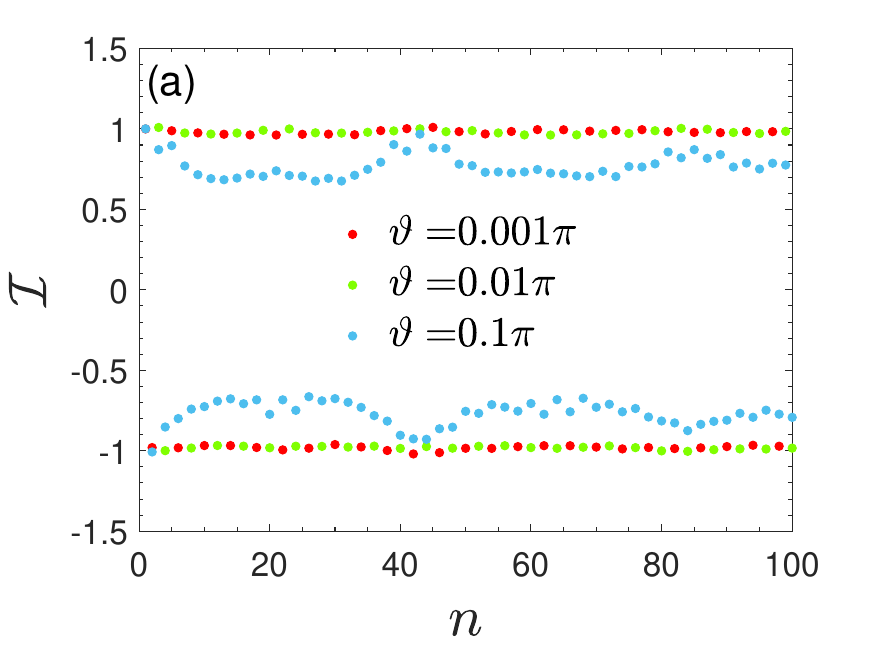}
    \includegraphics[width=0.49\linewidth]{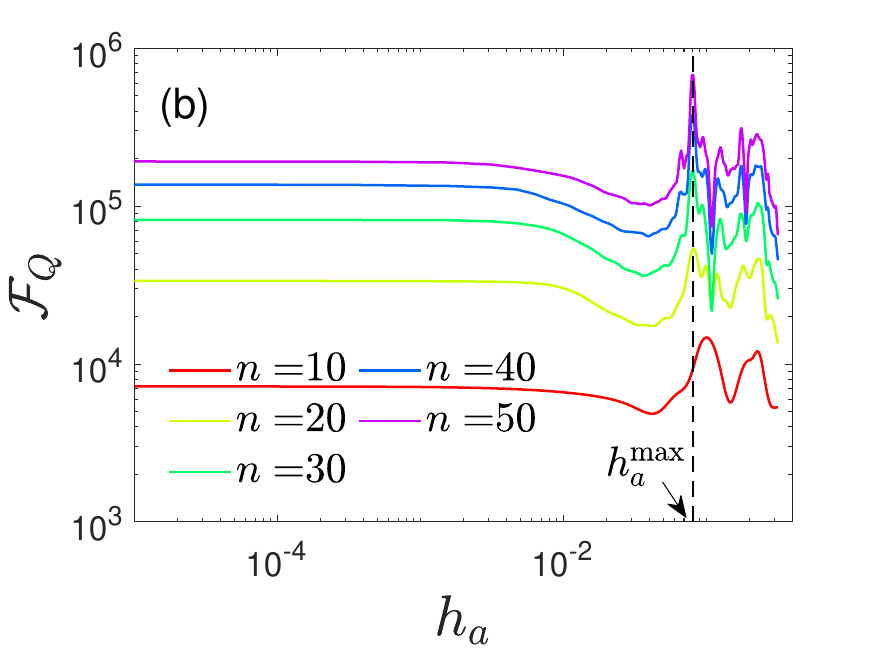}
    \caption{(a) The imbalance as a function of $n$ for different values of $\vartheta$ when $h_{a}{=}10^{-2}$. (b) The QFI versus $h_{a}$ over different stroboscopic times when $\vartheta{=}10^{-2}\pi$. In both panels $L{=}7$ and $\varepsilon{=}0.1$.   }
    \label{fig:IRS}
\end{figure}
Having analyzed the effect of off-resonance and crosstalk, we now explore the robustness of our DTC probe to imperfections in the initialization. 
Regardless of the presence or absence of the periodic-field, as a perfect quench with $\varepsilon{=}0$ preserves  total $z$-polarization in the system, the DTC order with arbitrary initialization is infinitely long-lived even in systems with finite sizes.
For $\varepsilon{>}0$, the configuration basis states of $H_I$ exhibit varying levels of robustness to imperfections, with some remaining resilient at higher values of $\varepsilon$.
In the previous section, we showed that initializing the system in state $|\psi(0)\rangle{=}|{\uparrow,\cdots,\uparrow}\rangle_{a}{\otimes}|{\downarrow,\cdots,\downarrow}\rangle_{b}$ provides persistence period-doubling oscillations of $\langle\mathcal{I}(nT)\rangle$ even for $\varepsilon{\neq}0$. 
Furthermore, significant overlap of this initial state with the "Schrödinger cat" state $|\Phi_{\pm}\rangle{=}\frac{1}{\sqrt{2}}(|\psi(0)\rangle {\pm}|\tilde{\psi}(0)\rangle)$ in the spectrum of the DTC has been identified as the main source of quantum-enhanced sensing. 
This obliges us to analyze the effect of perturbation in the initialization of the probe. For this, we consider a product state as 
\begin{align}\label{Eq.IS}
|\psi(0)\rangle &= \Pi_{j=1}^{L}|\varsigma_j\rangle_{a} \otimes |\varsigma_j\rangle_{b} \cr 
|\varsigma_j\rangle_{a} &= {\rm cos(\vartheta)} |\uparrow_j\rangle_{a} +   {\rm sin(\vartheta)} |\downarrow_j\rangle_{a} \cr   |\varsigma_j\rangle_{b} &= -{\rm sin(\vartheta)} |\uparrow_j\rangle_{b} +   {\rm cos(\vartheta)} |\downarrow_j\rangle_{b}
\end{align}
with $\vartheta{\in}[0,\pi/4]$, to avoid mixtures that yield $\langle\mathcal{I}(0)\rangle {=} 0$.
For the extreme values of $\vartheta {=} 0$, one can recover  $|\psi(0)\rangle$. However, for other values of $\vartheta$, a weighted mixture of different configurations is attainable.
In Fig.~\ref{fig:IRS}(a), we represent the normalized imbalance $\langle\mathcal{I}(nT)\rangle/\langle\mathcal{I}(0)\rangle$ over tens of period cycles when $\vartheta$ varies.
Clearly, the period-doubling oscillation of DTC remains unaffected by small imperfections during the initialization stage.
To assess the performance of our DTC probe, in Fig.~\ref{fig:IRS}(b), we plot the obtained QFI as a function of $h_a$ over different stroboscopic times when $\vartheta{=}10^{-2}\pi$.
The results show that this level of perturbation in the initial state has no tangible impact on the performance of our DTC probe.

\subsection{Imperfect quench effect}
\begin{figure}[t!]
    \centering
    \includegraphics[width=0.49\linewidth]{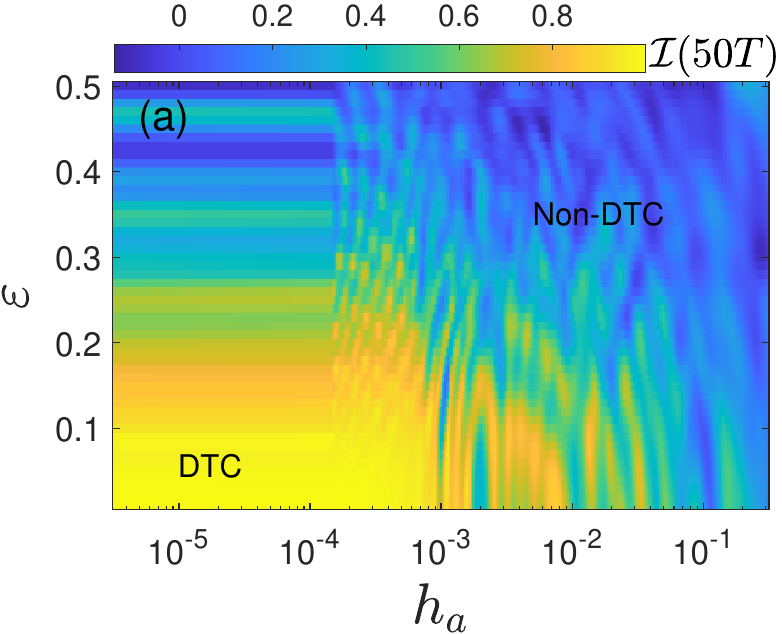}
    \includegraphics[width=0.49\linewidth]{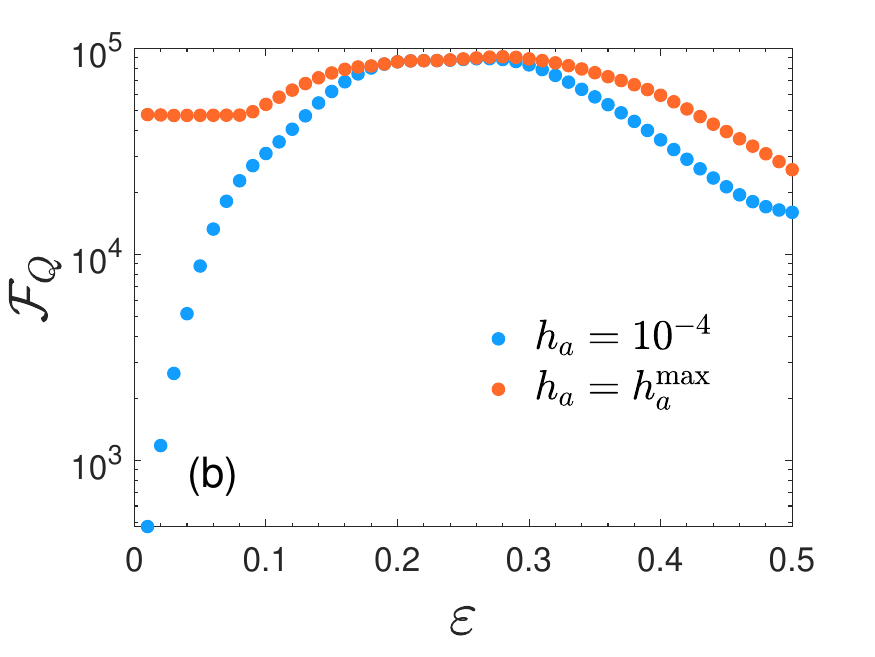}
    \caption{(a) The imbalance as a function of $h_a$ and $\varepsilon$ at $n{=}50$. 
    (b) The QFI versus $\varepsilon$ for different values of the $h_a$ at $n{=}20$. In both panels $L{=}7$.}
    \label{fig:PD}
\end{figure}
After explaining the sharp phase transition driven by $h_a$, we will now investigate the melting of the DTC by increasing the imperfection $\varepsilon$. 
In Fig.~\ref{fig:PD}(a), we plot the imbalance $\mathcal{I}(nT)$ after $n{=}50$ period cycles as a function of $h_a$ and $\varepsilon$. 
The phase diagram is entirely characterized by the parameters 
$h_a$ and $\varepsilon$. 
Increasing $\varepsilon$ reduces the DTC phase imbalance, but can improve the performance of the probe.
As evident in Fig.~\ref{fig:PD}(b), where we plot the obtained QFI as a function of $\varepsilon$ deep inside the DTC phase and at the transition point, by increasing the imperfection in the quench up to some value, the QFI grows. This growth in the DTC phase is more tangible than at the transition point. During the DTC phase, the dynamic is limited to a subspace spanned by $|\Phi_{\pm}\rangle$. Increasing $\varepsilon$ by engaging a larger portion of the Hilbert space encodes more information about $h_a$ in the quantum states.
This finding contrasts sharply with typical sensors, where imperfections diminish performance.

\section{Feasibility of quantum-enhanced sensitivity }
\begin{figure}[t!]
    \centering
    \includegraphics[width=0.49\linewidth]{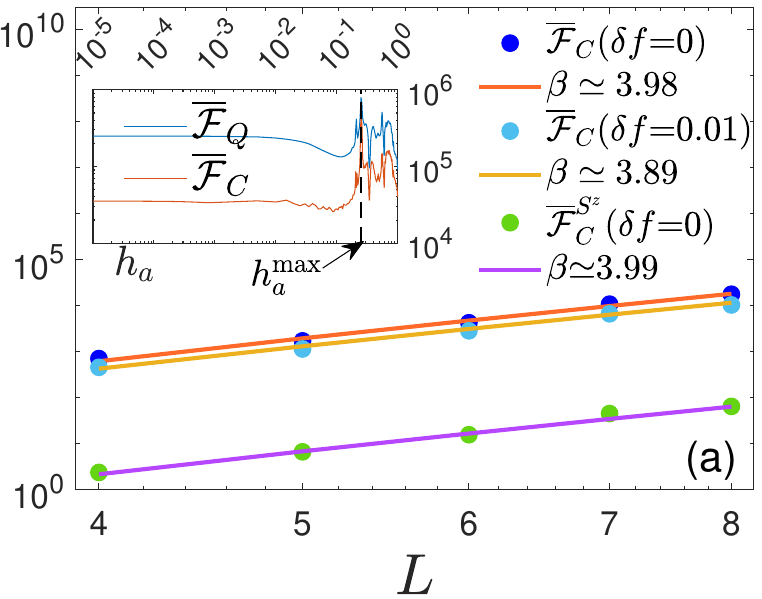}
    \includegraphics[width=0.49\linewidth]{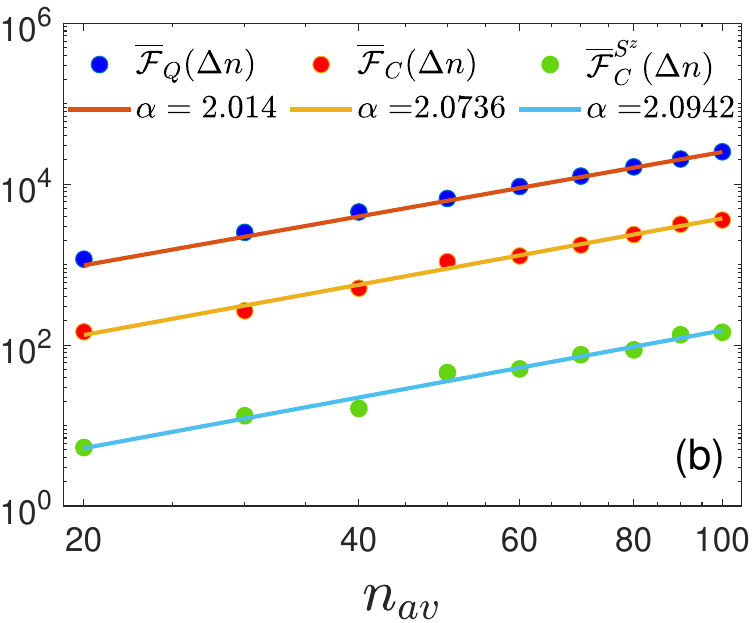}
    \caption{ (a) Time-averaged CFI over $N{=}50$ as a function of $L$ in DTC phase, namely $h_a{=}10^{-5}$ obtained using computational-basis (total magnetization) measurement, denoted by $\overline{\mathcal{F}}_C$ ($\overline{\mathcal{F}}^{S^{z}}_C$) for the probes in resonance ($\delta f{=}0$) and out of the resonance ($\delta f{=}0.01$) with the periodic-field. The markers show numerical results, and the solid lines represent the best-fit function as $\overline{\mathcal{F}}_C{\propto}L^{\beta}$ with reported $\beta$'s.
    Inset is $\overline{\mathcal{F}}_C$ and time-averaged QFI,  $\overline{\mathcal{F}}_Q$, over $N{=}100$ as a function of $h_a$ in a chain of size $L{=}7$ and $\delta f{=}0$. (b) Point averaged QFI and CFI versus $n_{av}$ in a system of size $L{=}7$ that operates in the DTC phase, namely $h_a{=}10^{-5}$. The markers are the numerical results, and the solid lines are the best fitting function as $\overline{\mathcal{F}}_{Q,C}{\propto} n_{av}^{\alpha}$ with reported $alpha$'s. All the results are obtained for $\varepsilon{=}0.1$.}
    \label{fig:CFI}
\end{figure}

Saturating the quantum Cram\'{e}r-Rao bound requires measuring the system using optimal measurements, which are generally complex and may even depend on the unknown parameter $h_a$. 
Therefore, finding an experimentally feasible set of measurements that results in $\mathcal{F}_{C}$ close to $\mathcal{F}_{Q}$ 
is highly desirable.
Interestingly, in our sensing scheme, simple measurements such as computational-basis measurement, described by $\{\Pi_{\mathbf{z}}{=}\ket{\mathbf{z}}\bra{\mathbf{z}}\}$ where in $\ket{\mathbf{z}}$ represents $2^{2L}$ different computational basis, or collective-spin observable of each chains described by $S^{\mu z}{=}\sum_{j=1}^{L}\sigma^{\mu z}_{j}$, 
can follow the scaling behavior of the QFI, concerning both time and size.
Here, we calculate the CFI for the computational-basis measurement using $p_{\mathbf{z}}{=}|\braket{\mathbf{z}}{\psi(n)}|^2$ and for the collective-spin observable using $p{=}\expectation{\psi(n)}{S^{a z}}{\psi(n)}$.
The results are denoted as $\overline{\mathcal{F}}_{C}$ and $\overline{\mathcal{F}}^{S^{z}}_{C}$, respectively.
The inset of Fig.~\ref{fig:CFI}(a) provides time-averaged CFI, defined as $\overline{\mathcal{F}}_C{=}\frac{1}{N}\sum_{n=1}^{N}\mathcal{F}_{C}$, and the time-averaged QFI as  $\overline{\mathcal{F}}_Q{=}\frac{1}{N}\sum_{n=1}^{N}\mathcal{F}_{Q}$. Here, $N$ is the total available time for sensing.
Similar to the QFI, the time-averaged CFI shows distinct behaviors in both the DTC phase and the non-DTC region. In addition, it peaks sharply at the transition point $h_a^{\max}$ and highly resembles the QFI. 
In Fig.~\ref{fig:CFI}(a), we show the scaling behavior of $\overline{\mathcal{F}}_C$, when the probe operates in the DTC phase, namely $h_a{=}10^{-5}$. Our results reveal that  $\overline{\mathcal{F}}_C$ follows the scaling behavior as $\overline{\mathcal{F}}_C{\propto} L^{\beta}$  with obtained $\beta{\simeq}3.98$ through fitting process.
Same scaling behavior is observed for $\overline{\mathcal{F}}^{S^{z}}_C{=}\frac{1}{N}\sum_{n=1}^{N}\mathcal{F}^{S^{z}}_{C}$, as can be seen from Fig.~\ref{fig:CFI}(a). In this case, we obtain $\beta{\simeq}3.99$. 
To extract the scaling behavior concerning time, we divide the total sensing time $N$ into $K$ smaller time intervals with equal space $n_{i+1}{-}n_i{=}\Delta n$ and average both QFI and CFI over these intervals. Fig.~\ref{fig:CFI}(b) represents the results as a function of $n_{av}{=}\sum_{i=1}^{K}(i\Delta n)$. 
While the markers are numerical results, the solid lines are the best fitting function as 
$\overline{\mathcal{F}}_{Q},\overline{\mathcal{F}}_{C},\overline{\mathcal{F}}^{S^{z}}_{C}{\propto} n_{av}^{\alpha}$ 
with $\alpha{=}2.014$ for the QFI, $\alpha{=}2.0736$ for the CFI using computational-basis measurement, and $\alpha{=}2.0942$ for the CFI using collective-spin measurement.  
These numerical analyses show that the simple measurement setups in our DTC probe can capture the quantum-enhanced sensitivity concerning both time and the size of the probe, represented by 
\begin{equation}
       \overline{\mathcal{F}}_{C}\propto L^{\beta}n^\alpha_{av} \quad \mathrm{with} \quad \beta{\simeq}4, \quad \alpha{\simeq}2. 
\end{equation}
This scaling behavior remains unchanged regardless of the frequency offset of the field. In Fig.~\ref{fig:CFI}(a), we plot the CFI using the computational-basis measurements as a function of $L$ for sensing the periodic-field with frequency offset  $\delta f{=}0.01$. The numerical results (markers) are properly described by the fitting function $\overline{\mathcal{F}}_{C}\propto L^{\beta}$ with $\beta{\simeq}3.89$.

\section{Effect of Decoherence }
\begin{figure}[t!]
    \centering
    \includegraphics[width=0.49\linewidth]{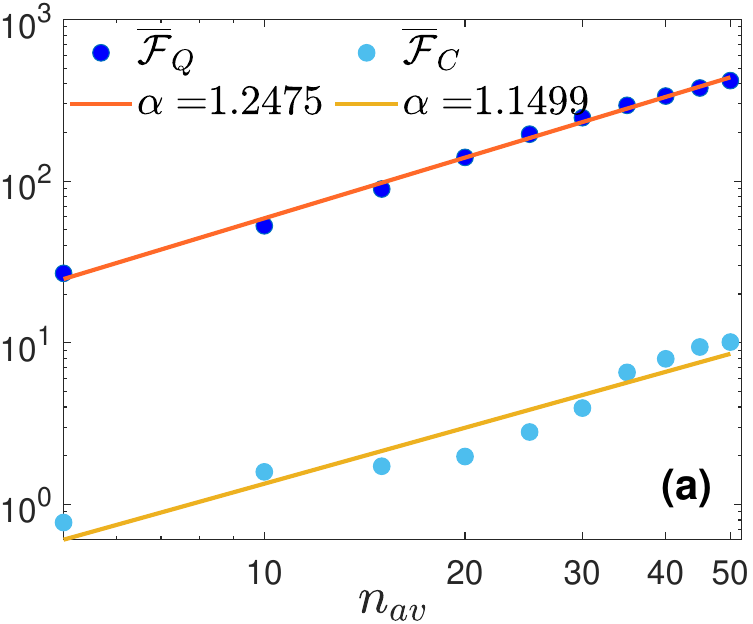}
    \includegraphics[width=0.49\linewidth]{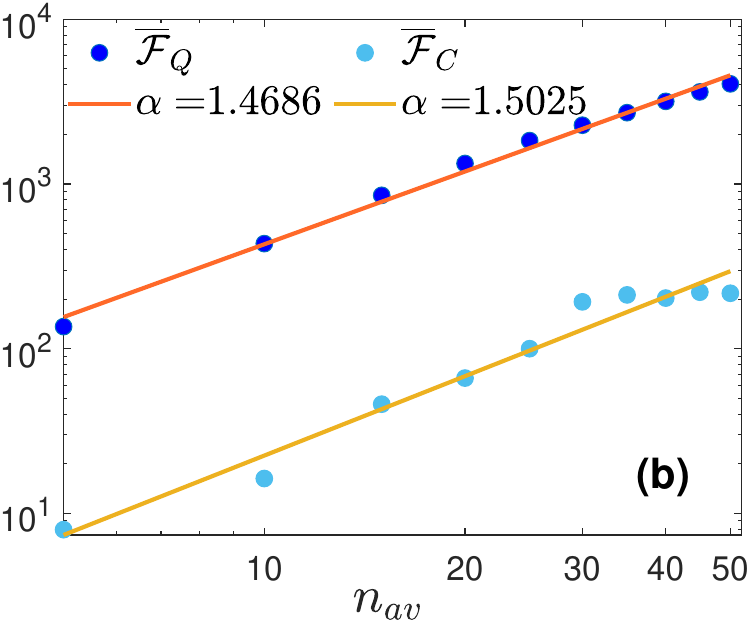}
    \caption{Point-averaged QFI and CFI versus $n_{av}$ in a system of size (a) $L{=}3$ and (b) $L{=}4$ that operates in the DTC phase, namely $h_a{=}10^{-5}$, in the presence of local dephasing with rate $\Gamma{=}10^{-3}$. The markers are the numerical results, and the solid lines are the best fitting function as $\overline{\mathcal{F}}_{Q,C}\propto n_{av}^{\alpha}$. All the results are obtained for $\varepsilon{=}0.1$.}
    \label{fig:Noise}
\end{figure}

Regardless of how much effort is put into isolation, quantum systems inevitably interact with their environment. These interactions can cause decoherence, dissipation, and energy exchange. Introducing an external environment, without fine-tuning, can disrupt subharmonic oscillations and thermalize the system~\cite{lazarides2017fate,yousefjani2025NHDTC,cenedese2025thermodynamics}. 
In this section, we analyze the effect of local dephasing on the sensing capability of the DTC probe. 
Considering Markovian diphasing noise, 
the system's density matrix evolves according to the time-dependent Lindblad equation as
\begin{widetext}
\begin{equation*}\label{Eq.Master}
\partial_t \rho =
\begin{cases}
-i[H_{a}+H_{b}+\Theta_n G^{az},\rho]+\Gamma \big(\mathcal{L}^{a}[\rho] + \mathcal{L}^{b}[\rho] \big) & (n-1)T<t\leqslant (n-\frac{1}{2})T \\
-i[H_{I}+\Theta_n G^{az},\rho]+\Gamma \big(\mathcal{L}^{a}[\rho] + \mathcal{L}^{b}[\rho] \big) & (n-\frac{1}{2})T<t\leqslant nT
\end{cases}  
\end{equation*}
\end{widetext}
here, $\mathcal{L}^{\mu}[\rho]{=}\sum_{j=1}^{L}\big( L_j^{\mu\dagger}\rho L^{\mu}_{j} -\frac{1}{2}L_j^{\mu\dagger}L^{\mu}_{j}\rho -\frac{1}{2}\rho L_j^{\mu\dagger}L^{\mu}_{j} \big)$ 
represents the  Liouvillian super-operator resulting from the coupling to the environment. 
$L^{\mu}_j{=}\sigma^{\mu z}_{j}$ is the Lindblad operator representing the dephasing process with rate $\Gamma$ in chain $\mu{=}a,b$.
To extract the scaling behavior with respect to time, in Fig.~\ref{fig:Noise}, we depict the point-averaged QFI and CFI as functions of $n_{av}$ in systems of size (a) $L{=}3$, and (b) $L{=}4$, when the probe operates in the DTC phase, namely $h_a{=}10^{-5}$, and exposed to the dephasing noise with rate $\Gamma{=}10^{-3}$.  
The behavior of numerical results (marker) can be described by the fitting function $\overline{\mathcal{F}}_{Q,C}\propto n_{av}^{\alpha}$ with the attached $\alpha$'s to the plots. 
Two main observations are in order. 
First, even with dephasing noise, a quantum enhancement in time, namely $\alpha>1$, is feasible. Second, as shown in Fig.~\ref{fig:Noise}, increasing the system size leads to a higher $\alpha$'s, resulting in a greater quantum enhancement. 

\section{Experimental proposals}
The methodology presented here is independent of the underlying physics, making it applicable to a broad spectrum of quantum and even classical systems. 
Here, we provide a proof-of-principle demonstration of our DTC sensor in ultracold atoms on an optical lattice, which offers a promising platform for experimental realization of the proposed DTC due to its broad controllability and flexibility. Specifically, the precisely tunable height and width of the lasers used to create the optical trapping potential enable the demanding quench required by our protocol. 
A detailed discussion on implementing this DTC using dipolar gases in an optical lattice can be found in Ref.~\cite{huang2018clean}. 
Using the Jordan-Wigner transformation, one can easily map the Hamiltonian Eq.~(\ref{Eq.Hamiltonian}) to the spinless fermion model as
\begin{equation}\label{Eq.FermiHamiltonian}
H(t) = 
\begin{cases}
H_a{+}H_b={-}J_z\sum_{j=1}^{L-1}\big(n^{a}_{j}n^{a}_{j+1} + n^{b}_{j}n^{b}_{j+1}\big) & 0{<}t{\leqslant} t_1 \\
H_{I} = J_{ab}\sum_{j=1}^{L}( c^{a\dagger}_j c^b_{j} + c^{a}_j c^{b\dagger}_{j}) & t_1{<}t{\leqslant} t_1 {+} t_2,
\end{cases}  
\end{equation}
where $n^{\mu}_{j}{=}c^{\mu\dagger}_jc^{\mu}_j$ counts the particle numbers at site $j$ in chain $\mu{=}a,b$, and 
$c^{\mu\dagger}_j$ ($c^{\mu}_j$) is the fermionic creation (annihilation) operator.  
In this notation, the periodic magnetic field can be rewritten as 
\begin{equation}
 H_{ac}(t){=}B_a(t)\sum_{j=1}^{L}j n^{a}_{j}.
\end{equation}
 Certain fermionic atoms, such as dysprosium ($^{163}$Dy) and erbium ($^{167}$Er) exhibit significant magnetic dipole moments as ${\approx}10\mu_B$ and ${\approx}7\mu_B$, respectively. Here, $\mu_B$ is the Bohr magneton.
When cooled to ultracold temperatures and loaded into an optical lattice, these atoms can demonstrate behaviors governed by magnetic dipolar interactions.
The magnetic dipole-dipole interactions, which are long-range and anisotropic, take the following form
\begin{equation}
    V_{jj^{'}}=\frac{C_{dd}}{|r_{jj^{'}}|^3}\Big(1-3(\hat{d}_{j}.\hat{r}_{jj^{'}})(\hat{d}_{j^{'}}.\hat{r}_{jj^{'}}) \Big)
\end{equation}
where $C_{dd}$ is the dipole-dipole interaction strength, 
$\hat{r}_{jj^{'}} {=}\vec{r}_j -\vec{r}_{j^{'}}/|\vec{r}_{j} -\vec{r}_{j^{'}}|$, and $r_{jj^{'}}{=}|\vec{r}_j -\vec{r}_{j^{'}}|$. 
The unit vector $\hat{d}_j$ denotes the direction of the magnetic dipole of an atom at the site $j$, with coordinate $\vec{r}_j{=}x_j\hat{x}{+}y_j\hat{y}$, which is usually controlled by applying a uniform magnetic field.
Implementing our DTC system requires an engineered optical potential to form a 1D lattice as well as a double-well potential with controllable barriers.   
During $t_1$, one can suppress the interaction between two chains $a$ and $b$ by ramping up the barrier of the double-well. Therefore, the dipolar interaction between the atoms with distance $x$ in each chain, with strength $C_{dd}/x{|j_\mu{-}j^{'}_{\mu}|}$, generates $H_{a}{+}H_{b}$ in Eq.~(\ref{Eq.Hamiltonian}). 
Despite its long-range nature, its decay to the third power with distance likely does not impact the main results.
To implement the tunneling between two chains $a$ and $b$, one needs to lower the barrier of the double-well, increase the optical potential to suppress the intra-chain interaction, and apply a uniform magnetic field as $\vec{B}_2{=}\sqrt{1/3}\hat{x}{+}\sqrt{2/3}\hat{z}$, that results in 
\begin{align}\label{Eq.Exp1}
 V_{j_a{\neq}j^{'}_a}&{=}0,   \quad
 V_{j_b{\neq}j^{'}_b}{=}0,    \quad
 V_{j_a{=}j_b} {=} \frac{C_{dd}}{y^3} \cr
 \cr
 V_{j_a{\neq}j_b}&{=} \frac{C_{dd}}{r_{ab}^3}(1-\frac{|j_a-j_b|^2 x^2}{r_{ab}^2}),
\end{align}
where $r_{ab}{=}\sqrt{y^2+|j_a-j_b|^2x^2}$.
Setting $y{\gg} x$, leads to $V_{j_a{=}j_b}{\rightarrow}0$ and $V_{j_a{\neq}j_b}{\rightarrow}0$ and allows tunneling for   
duration $t_2{\simeq}\hbar x^3{/}C_{dd}$.
\\ \\
Imposing this probe on a periodic-field as $\vec{B}_a(nt){=}(-1)^{n}h_a\hat{y}$, while may not affect the intra-chain dipole-dipole interaction, results in 
\begin{align}\label{Eq.Exp2}
 V_{j_a{\neq}j^{'}_a}&{=}0,    \quad V_{j_b{\neq}j^{'}_b}{=}0, \quad   V_{j_a{=}j_b} {=} \frac{C_{dd}}{y^3}(1{-}\frac{x^2}{y^2}(1{+}\sqrt{3}(-1)^n j_a h_a\frac{y}{x}))\cr
 V_{j_a{\neq}j_b}&{=} \frac{C_{dd}}{r_{ab}^3}(1-\frac{(1 +\sqrt{3}\frac{(-1)^nj_a h_a y}{|j_a-j_b|x})|j_a-j_b|^2 x^2}{r_{ab}^2}).
\end{align}
 This result shows that, for infinitesimal $h$ and $y{\gg}x$, one can recover Eq.~(\ref{Eq.Exp1}) that allows tunneling due to the suppression of the dipole-dipole interaction. However, larger values of $h$ through inducing site-dependent dipole-dipole interactions affect the tunneling between chains and, hence, demolish the DTC nature.
For the lattice spacing reported in a typical experiment $x {\simeq} 266~\mathrm{nm}$ (chain spacing $y {\simeq} 10x$), the nearest-neighbour dipolar coupling produces an interaction energy  $V_{j_{a}{=}j_{b}}/(2\pi\hbar)\simeq 60\mathrm{Hz}$ for Dy and $V_{j_{a}{=}j_{b}}/(2\pi\hbar)\simeq29.4\mathrm{Hz}$ for Er. Using $t_2\simeq\hbar x^{3}/C_{dd}$ one obtains $t_2\simeq1/(2\pi\cdot60)\approx2.65~\mathrm{ms}$ for Dy  and  $\simeq1/(2\pi\cdot29.4)\approx5.41~\mathrm{ms}$ for Er. Adopting the binary-quench timing used here ($t_{1}=t_{2}=1/(2J_{z})$ with the energy scale chosen so that $J_{z}/(2\pi)\sim f_{j_a{=}j_b}$), the drive period is $T\sim2.65~\mathrm{ms}$ (Dy) and $\sim5.41~\mathrm{ms}$ (Er), so a single-cycle repetition rate is $1/T\sim377~\mathrm{Hz}$ (Dy) and $\sim185~\mathrm{Hz}$ (Er). 
For typical coherence time $T_{2}=0.1\ \mathrm{s}$ in the dipolar systems one has  $\left\lfloor\frac{100}{2.65}\right\rfloor = 37$ for Dy and $\left\lfloor\frac{100}{5.41}\right\rfloor = 18$ for Er of cycles possible. Therefore, using a DTC sensor that includes $L$ atoms can result in $0.0270/L^2 ~pT/m/\sqrt{\mathrm{Hz}}$ and $0.0555/L^2 ~pT/m/\sqrt{\mathrm{Hz}}$ using Dy and Er, respectively.

\section{Conclusion}
In this paper, we propose a DTC in a disorder-free system for sensing a gradient periodic-field with quantum-enhanced sensitivity. 
Our results show that a many-body probe initialized in a product state can achieve ultimate precision, namely $\mathcal{F}_{Q}{\propto}n^2L^4$, provided that it operates in the DTC phase. 
The $L^4$ scaling of the QFI is due to the resonance dynamics between two GHZ-type eigenstates in the DTC regime, which amplify the response of the system. The coherent accumulation over $n$ Floquet cycles leads to the observed $n^2$ time scaling, and the resulting sensing bandwidth is centered at the drive resonance and limited by the coherence time $T_2$ of the system. This enhancement is experimentally achievable through measuring the standard projective measurement.
Remarkably, this quantum enhancement shows robustness against different imperfections and dephasing noise in the sensing protocol. 
Our results show that the DTC sensor that operates in the DTC phase for $n{\leq}50$ periodic quench keeps the behavior of the QFI 
for a periodic-filed with frequency offset $\delta f$ of the DTC's frequency when $\delta f n{\leq}1{/}2$, for a crosstalk with strength $\eta{<}1$, for an initialization with perturbation parameter $\vartheta{\leq}10^{-2}\pi$, for an imperfect quench of strength $\varepsilon{<}0.2$, and finally for a local dephasing noise with rate $\Gamma{\leq}10^{-3}$.   
In addition, we discuss the implementation of our probe in an optical lattice setup. 
Our results open avenues for harnessing inherent entanglement in interacting, non-equilibrium quantum many-body systems for enhanced metrology.

\section{Acknowledgment}
AB acknowledges support from
National Natural Science Foundation of China (Grants
Nos. $12050410253$, $92065115$, and $12274059$).

% \bibliography{Ref}

\end{document}